# Why Professor Richard Feynman was upset solving the Laplace equation for spherical waves?

Anzor A. Khelashvili[a]

Institute of High Energy Physics, Iv. Javakhishvili Tbilisi State University, University Str. 9, 0109, Tbilisi, Georgia and St. Andrea the First-called Georgian University of Patriarchy of Georgia, Chavchavadze Ave. 53a, 0162, Tbilisi, Georgia

Teimuraz P. Nadareishvili[b]

Iv. Javakhishvili Tbilisi State University, Faculty of Exact and Natural Sciences, Chavchavadze Ave. 3, 0179, Tbilisi, Georgia and Institute of High Energy Physics, Iv. Javakhishvili Tbilisi State University, University Str. 9, 0109, Tbilisi, Georgia

**Abstract.** We take attention to the singular behavior of the Laplace operator in spherical coordinates, which was established in our earlier work. This singularity has many non-trivial consequences. In this article we consider only the simplest ones, which are connected to the solution of Laplace equation in Feynman classical books and Lectures. Feynman was upset looking in his derived solutions, which have a fictitious singular behavior at the origin. We show how these inconsistencies can be avoided.

## I.     INTRODUCTION

R.Feynman in his "Lectures"[1] discussed the derivation of spherical waves on the basis of wave equation in spherical coordinates. Derived solution has a singularity at the origin $r = 0$. He wrote about this solution the following: "Our solution must represent physically a situation, in which there is some source at the origin. In other words, we have inadvertently made a mistake. We have not solved the free wave equation *everywhere;* we solved it with zero on the right everywhere except at the origin. Our mistake crept in because some of the steps in our derivation are not "legal" when $r=0$.".

R. Feynman probably had meant singular behavior of Laplace operator at the origin. Nowadays we know that Laplacian is indeed singular in spherical coordinates[2,3] and a more caution treatment is necessary.

Our aim in this article will be careful investigation of Feynman's problem. This article is organized as follows. In Sec. II we consider the electrostatic problem. In Sec. III we consider the Yukawa potential. Sec. IV we give concluding remarks.

## II. ElECTROSTATIC PROBLEM

Let us begin, follow Feynman, by electrostatic problem, where the same mistake occurs. R.Feynman mentioned**:** "Let's show that it is easy to make the same kind of mistake in an electrostatic problem. Suppose we want a solution of the equation for an electrostatic potential in free space, $\nabla^2 \psi = 0$''.

In an explicit form this equation looks like

$$\nabla^2 \psi(r) = \frac{d^2 \psi}{dr^2} + \frac{2}{r}\frac{d\psi}{dr} = 0 \qquad (1)$$

R.Feynman continued**:** "It is often more convenient to write this equation in the following form

$$\nabla^2 \psi = \frac{1}{r}\frac{d^2}{dr^2}(r\psi) \qquad (2)$$

If you carry out the differentiation indicated in this equation, you will see that the right hand side is the same as in previous equation"

We want to emphasis that exactly this statement fails at $r = 0$ [4]. It was shown in [4] that the correct relation looks like

$$\nabla^2 \psi = \frac{1}{r}\frac{d^2}{dr^2}(r\psi) - 4\pi\delta^{(3)}(r)(r\psi) \tag{3}$$

Therefore some of relations of Feynman's book will undergo relevant corrections. If we introduce the representation of 3-dimensional delta function in spherical coordinates, namely,

$$\delta^{(3)}(r) = \frac{\delta(r)}{4\pi r^2}, \tag{4}$$

and use traditional short relation

$$u(r) = r\psi(r) \tag{5}$$

we derive the following form of Laplace equation

$$r\frac{d^2 u}{dr^2} - \delta(r)u(r) = 0 \tag{6}$$

It seems that after transition to function $u(r)$ there appears source-like term in the Laplace equation $\delta(r)u(r) = \delta(r)u(0)$. It is caused by singular character of Jacobian of transformation from Cartesian to spherical coordinates, $J = r^2 \sin\theta$ at the origin. (usually the singularity with respect to $\theta$ is avoided by the requirements of discontinuity and uniqueness, which results in appearance of spherical harmonics $Y_{lm}(\theta,\varphi)$ [5].

How can we reject this "extra" term from equation? It depends on value of $u(0)$. there are three possibilities: a finite $u(0)$ must be excluded, because after returning to $\psi$, there appears undesirable $1/r$ term, which is not a solution of Laplace equation. The second possibility $u(0) = \infty$ must also be rejected, as the presence of infinite term in equation is nonsense. So, there remains only possibility, $u(0)=0$ and if at the same time we take this function to behave as

$$u(r) \underset{r \to 0}{\approx} r^{1+\varepsilon}, \quad (\varepsilon > 0) \tag{7}$$

then this term kills the delta function and we return to the standard equation.

This consideration shows, that the standard reduced radial equation is valid only in cases when the restriction $u(0)=0$ is fulfilled[3].

In this case standard equation looks like

$$\frac{d^2 u}{dr^2} = 0 \tag{8}$$

and has a solution

$$u(r) = ar + b \tag{9}$$

Taking into account the above restriction, we obtain that $b = 0$ and, therefore $\psi = \frac{u}{r} = a = const.$

While R.Feynman mentioned: "… found that the following $\psi$ is a solution for the electrostatic potential in free space

$$\psi = a + \frac{b}{r} \tag{10}$$

Then R.Feynman continued:" Something is evidently wrong. In the region where there are no electric charges, we know the solution for the electrostatic potential: the potential is everywhere constant. That corresponds to the first term in our solution. But we have also the second term, which says that there is a contribution to the potential that varies as one over the distance from the origin. We know, however, that such a potential corresponds to a point charge at the origin. So, although we thought we were solving for the potential in free space, our solution also gives the field for a point source at the origin".

We see that rigorous consideration of radial Laplace equation puts all things on their own places. The restriction $u(0) = 0$ has a decisive meaning. The natural question arises: Why does the "boundary condition-like" restriction arises in the free equation? This happens because transformation to spherical coordinates does not involve $r = 0$ point and the transition to $u(r)$ function feels this, because of $r^{-1}$ factor.

The second term in Feynman consideration $b/r$ is not a solution at all. Indeed, after its substitution into Laplace equation we obtain a delta function, instead of zero. Another way to avoid this solution is a comparison to Cartesian solution, where the wave function at the origin is constant, as it is clear from the solution of the $\nabla^2 \psi = 0$ equation in this coordinates[6]

$$\psi = e^{\pm i\alpha x} e^{\pm i\beta y} e^{\pm \sqrt{\alpha^2 + \beta^2} z} \quad (11)$$

The same can be demonstrated by considering characteristic equation for (1), substituting there $\psi \approx r^s$, it follows

$$s(s+1)r^{s-2} = 0 \quad (12)$$

integrating this equation by the spherical element $r^2 dr$ in arbitrary bounds, we obtain

$$s(s+1)\frac{1}{s+1}\left(r^{s+1}\right)_a^b = 0 \quad (13)$$

it follows that we have only one solution $s = 0$, which corresponds to $\psi = const$, in accordance with Cartesian behavior.

## III. YUKAWA POTENTIAL

Another place where R.Feynman made use the relation (2) is the Yukawa potential (ibid. Chapter 28). It is commonly believed that the Yukawa potential is a spherically symmetric solution of well-known wave equation

$$\nabla^2 \phi - \mu^2 \phi = 0 \tag{14}$$

Arming with the previous consideration, R.Feynman wrote this equation in the following form

$$\frac{1}{r}\frac{\partial^2}{\partial r^2}(r\phi) - \mu^2 \phi = 0 \tag{15}$$

solution of which is $\phi = K e^{\pm \mu r}$ and therefore after a suitable boundary condition at infinity, it follows the Yukawa potential

$$\phi = K \frac{e^{-\mu r}}{r} \tag{16}$$

But we know, that a rigorous application of correct relation (3) gives

$$\nabla^2 \frac{e^{-\mu r}}{r} = \mu^2 \frac{e^{-\mu r}}{r} - 4\pi \delta^{(3)}(r) e^{-\mu r} \quad , \tag{17}$$

Interesting enough that this form was given in the earlier book[7].

It follows that the Yukawa potential is not a spherically symmetric solution of Eq. (14) *everywhere*, but it is such only ahead of origin.

We see from Eq. (17) that the Yukawa potential is a solution of nonhomogenious wave equation with a sourse term on the right-hand side

$$\nabla^2 \phi - \mu^2 \phi = -4\pi K \delta^{(3)}(r) \tag{18}$$

## IV. CONCLUSIONS

In conclusion, we have demonstrated that careful consideration of Laplacian operator near the origin removes all the inconsistencies related to Feynman's analysis. The only point which must be clarified is a rigorous determination of a character of aspiration to zero of $u(r)$ function in order to provide $u(r)\delta(r) = 0$, which is a business of the theory of distributions.

Moreover we note that things will be changed in many applications where the Laplace operator is used in spherical coordinates.

## ACKNOWLEDGEMENTS

Authors acknowledges financial support of the Shota Rustaveli National Science Foundation (Projects DI/13/02 and FR/11/24)

---

[a] Electronic mail: anzor.khelashvili@tsu.ge
[b] Electronic mail: teimuraz.nadareishvili@tsu.ge